\documentstyle[12pt]{article}
\newlength{\extraspace}
\newlength{\extraspaces}
\newcommand{\be}{\begin{equation}
\addtolength{\abovedisplayskip}{\extraspaces}
\addtolength{\belowdisplayskip}{\extraspaces}
\addtolength{\abovedisplayshortskip}{\extraspace}
\addtolength{\belowdisplayshortskip}{\extraspace}}
\newcommand{\ee}{\end{equation}}
\newcommand{\ba}{\begin{eqnarray}
\addtolength{\abovedisplayskip}{\extraspaces}
\addtolength{\belowdisplayskip}{\extraspaces}
\addtolength{\abovedisplayshortskip}{\extraspace}
\addtolength{\belowdisplayshortskip}{\extraspace}}
\newcommand{\ea}{\end{eqnarray}}
\makeatletter

\@addtoreset{equation}{section}
\makeatother
\newcommand{\nonu}{\nonumber \\[.5mm]}
\newcommand{\A}{&\!\!\!}
\begin{document}
\thispagestyle{empty}
\setlength{\baselineskip}{6mm}
\begin{flushright}
SIT-LP-09/06 \\
%{\tt arXiv:yymm.nnnn[hep-th]} \\
June, 2009
\end{flushright}
\vspace{3mm}
\begin{center}
{\large{\bf On gauge coupling constant \\[2mm]
in linearization of nonlinear supersymmetry}} \\[18mm]
{\sc Kazunari Shima}
\footnote{
\tt e-mail: shima@sit.ac.jp} \ 
and \ 
{\sc Motomu Tsuda}
\footnote{
\tt e-mail: tsuda@sit.ac.jp} 
\\[3mm]
{\it Laboratory of Physics, 
Saitama Institute of Technology \\
Fukaya, Saitama 369-0293, Japan} \\[18mm]
\begin{abstract}
We study in two space-time dimensions ($d = 2$) the relation between $N = 2$ supersymmetric (SUSY) QED theory 
and $N = 2$ nonlinear (NL) SUSY model by linearizing $N = 2$ NLSUSY generally 
based upon the fundamental notions of the basic theory. 
We find a remarkable mechanism which determines theoretically 
the magnitude of the bare gauge coupling constant from the general structure of auxiliary fields. 
We show explicitly in $d = 2$ that 
the NL/linear SUSY relation (i.e. a SUSY compositeness condition for all particles) 
determines the magnitude of the bare electromagnetic coupling constant (i.e. the fine structure constant) 
of $N = 2$ SUSY QED. 
\\[5mm]
\end{abstract}
\end{center}

\newpage

\section{Introduction}

\noindent
Towards the unified description of space-time, all forces and matter beyond the standard model (SM) 
it is promising and essential to study (new) physics based upon super-Poincar\'e (SP) symmetry. 

From the group theoretical survey of $SO(N)$ SP algebra for $N > 8$ \cite{KS1}, 
it was shown that $SO(10)$ SP group accomodates minimally (and correctly) 
the SM with just three generations of quarks and leptons 
and the graviton in the {\it single} irreducible representation. 
According to $SO(10) \supset SU(5) \supset SU(3) \times SU(2) \times U(1)$ 
and the decomposition of 10 supercharges 
$\underline{10}_{SO(10)} = \underline{5}_{SU(5)} + \underline{5}^*_{SU(5)}$ 
we regarded $\underline{5}$ as a hypothetical spin-${1 \over 2}$ new particles (called {\it superons}) 
possessing the same quantum numbers as $\underline{5}$ of $SU(5)$ GUT 
and proposed superon-quintet model (SQM) \cite{KS2} for matter and forces. 
%
%Typical processes in particle physics (up to those for the new physics beyond the SM) 
%were exemplified in terms of multiple (composite) diagrams of 
%more fundamental (hypothetical) spin-1/2 objects {\it superon} 
%which correspond to $\underline{5}$ of $SU(5)$ 
%(with the same quantum numbers as those of $SU(5)$ GUT) in the decomposition of 10 supercharges as 
%$\underline{10}_{SO(10)} = \underline{5}_{SU(5)} + \underline{5}^*_{SU(5)}$ 
%according to $SO(10) \supset SU(5)$ ({\it superon-quintet model} (SQM)) \cite{KS2}. 
%

For the field theoretical arguments of SQM including gravity, called {\it superon-graviton model} (SGM), 
a fundamental action has been constructed in nonlinear supersymmetric general relativity (NLSUSY GR) theory \cite{KS3}, 
which is based upon the general relativity (GR) principle 
and the nonlinear (NL) representation \cite{VA} of supersymmetry (SUSY) \cite{WZ,GL}. 
%
%by inspiring spinor supercurrents \cite{KS4} of the ($N = 1$) NLSUSY Volkov-Akulov (VA) model \cite{VA} 
%giving the supercharges (superon) in terms of Nambu-Goldstone (NG) fermions. 
Also, the NLSUSY GR action has a priori promising large symmetries 
isomorphic to $SO(N)$ ($SO(10)$) SP group \cite{ST1}. 

The NLSUSY Volkov-Akulov (VA) model \cite{VA} is an almost {\it unique action} 
which represents $N = 1$ global SUSY nonlinearly and describes the dynamics of massless Nambu-Goldstone (NG) fermion 
of the spontaneous breakdown of space-time SUSY, i.e. the spontaneous SUSY breaking (SSB) mechanism 
is encoded {\it a priori} in the geometry of ultimate flat space-time. 
(Alternatively, the geometry of ultimate flat (tangential) space-time is defined by the invariant volume action 
and induces self-contained phase transition to ordinary Minkowski space-time accompanying NG fermion, 
whose invariant action is given by the NLSUSY VA model.) 
While, the NLSUSY model is recasted (related) to various linear (L) SUSY theories 
with SSB (abbreviated as {\it NL/L SUSY relation}) by linearizing NLSUSY, 
which has been shown by many authors in the various cases both for free theories \cite{IK}-\cite{lin-ST2} 
and for interacting (Yukawa interactions and SUSY QED) theories \cite{lin-ST3}-\cite{lin-ST4b}. 
(In Ref.\cite{IK} SUSY QED theory in NL/L SUSY relation is discussed by using a constrained 
gauge superfield in the context of the coupling of the VA action 
to the gauge multiplet action.) 

From the above group and field theoretical aspects in SGM, 
it is interesting and worthwhile investigating the NLSUSY GR theory as a fundamental theory beyond the SM, 
which proposes a new paradigm, SGM scenario \cite{KS2,KS3,ST1,ST2}, for describing the unity of nature. 

Indeed, the NLSUSY GR action is constructed in the Einstein-Hilbert form on new (generalized) space-time, 
{\it SGM space-time} \cite{KS3}, where tangent space-time has the NLSUSY structure, 
i.e. flat space-time is specified not only by the $SO(3,1)$ Minkowski coodinates 
but also by $SL(2,C)$ Grassmann coordinates for NLSUSY as coset parameters of ${super GL(4,R) \over GL(4,R)}$. 
The locally homomorphic non-compact groups $SO(3,1)$ and $SL(2,C)$ 
for the new (SGM) space-time degrees of freedom 
is analogous to compact groups $SO(3)$ and $SU(2)$ for gauge degrees of freedom (d.o.f.) of 't Hooft-Polyakov monopole. 
The cosmological term (for the cosmological constant $\Lambda > 0$) in the NLSUSY GR action 
can be interpreted as the potential which induces 
the SSB of super-$GL(4,R)$ down to $GL(4,R)$ due to the NLSUSY stucture, 
i.e. the phase transion of SGM spacetime to Riemann spacetime with NG-fermion superons matter, 
called {\it Big Decay} \cite{ST2} which subsequently ignites the Big Bang and the inflation of the present universe. 
Since the cosmological term gives the NLSUSY model \cite{VA} 
in {\it asymptotic} Riemann-flat (i.e. the ordinary vierbein $e^a{}_\mu \rightarrow \delta^a{}_\mu$) space-time, 
the SSB scale of NLSUSY, arbitrary so far, is now related to 
the Newton gravitational constant $G$ and the cosmological term of GR. 

All (observed) particles, which are assigned uniquely 
into the single irreducible representation of $SO(N)$ ($SO(10)$) SP group 
as an on-shell supermultiplet of $N$ LSUSY, are considered to be realized 
as (massless) eigenstates of $SO(N)$ ($SO(10)$) SP composed of $N$ (spin-${1 \over 2}$) 
massless NG-fermion superons through the NL/L SUSY relation after Big Decay. 
Note that the no-go theorem is overcome (circumvented) in a sense that 
the nontivial $N$-extended SUSY gravity theory with $N > 8$ has been constructed in the NLSUSY invariant way, 
i.e. by the degenerate vacuum (flat space-time). 

As for the low energy physics of NLSUSY GR in the asymptotic Riemann-flat space-time, 
we showed that the phase transition from the massless superon-graviton (SGM) phase to 
the composite eigenstates-graviton phase corresponds to the transition to the 
true vacuum (the minimum of the potential) \cite{ST3,STL}. 
In those works the SSB scale induces (naturally) a fundamental mass scale originated from the cosmological constant 
and gives through the NL/L SUSY relation a simple explanation 
of the mysterious (observed) numerical relation 
between the (four dimensional) dark energy density of the universe and the neutrino mass \cite{ST2,ST3} 
in the vacuum of the $N = 2$ SUSY QED theory 
(in two-dimensional space-time ($d = 2$) for simplicity) \cite{STL}. 
(Note that the $N = 2$ SUSY gives the minimal and realistic model \cite{STT} in the SGM scenario.) 
These are particle physics consequences of NLSUSY GR, i.e. the relation between the large scale structure 
of space-time and the low energy particle physics. 

In the above SGM scenario (the Big Decay process from NLSUSY GR), 
it is also interesting and crucial problem how the (bare) gauge coupling constant of the SUSY QED theory 
is determined or whether it is calculable or not, provided the fundamental theory is that of everything. 
We focus on this problem in this paper by studying the NL/L SUSY relation for the $N = 2$ SUSY QED theory in $d = 2$ 
with a {\it general} structure of auxiliary fields from more general scheme \cite{STgc1,STgc2}. 
The relation between the $N = 2$ NLSUSY model and the $N = 2$ SUSY QED ($U(1)$ gauge) theory 
has been shown in $d = 2$ \cite{lin-ST3,lin-ST4b} 
under the adoption of the simplest {\it SUSY invariant constraints} 
and the subsequent {\it SUSY invariant relations} \cite{lin-ST2,lin-ST4a}. 
The SUSY invariant relations, which are obtained {\it systematically} from the SUSY invariant constraints 
\cite{IK,UZ,lin-ST2} in the superfield formulation \cite{WB}, describe all component fields 
in the LSUSY multiplets as the composite eigenstates in terms of the NG-fermion superons. 

In this paper we study for simplicity and without loss of generality 
the $N = 2$ SUSY QED, which is physically minimal case, theory in $d = 2$ by linearizing $N = 2$ NLSUSY 
under {\it general} SUSY invariant constraints which induce the subsequent {\it general} SUSY invariant relations. 
In the general NL/L SUSY relation (i.e. the over-all SUSY compositness condition) 
we find a remarkable mechanism which determines the magnitude of the bare gauge coupling constant 
from vacuum expectation values (vevs) (constant terms in SUSY invariant relations) 
of auxiliary scalar fields. 
We show explicitly in $d = 2$ that the NL/L SUSY relation determines 
the magnitude of the bare electromagnetic coupling constant 
(i.e. the fine structure constant) of $N = 2$ SUSY QED. 

This paper is organized as follows. 
In Section 2 we present a $N = 2$ general SUSY QED gauge action in $d = 2$ 
both in the superfield formulation and in the component form. 
In Section 3 we show in the linearization of $N = 2$ NLSUSY in $d = 2$ 
the SUSY invariant constraints and the SUSY invariant relations in the most general form. 
After some reductions of those constraints and relations to more simple (but nontrivial and general) 
expressions in Section 3 (for simplicity of arguments), 
the relation between the $N = 2$ NLSUSY action and the $N = 2$ SUSY QED ($U(1)$ gauge) action 
is discussed in Section 4, where the gauge coupling constant depends on the vevs 
(constant terms in SUSY invariant relations) of auxiliary scalar fields. 
Summary and discussion are given in Section 5.

\section{$N = 2$ SUSY QED action in $d = 2$}

\noindent
Let us first introduce the superfield formulation of $N = 2$ SUSY QED theory in $d = 2$, 
in which a $N = 2$ general SUSY QED action is constructed from $N = 2$ general gauge 
and $N = 2$ scalar matter superfields on superspace coordinates $(x^a, \theta_\alpha^i)$ ($i = 1, 2$). 
The $d = 2$, $N = 2$ general gauge superfield \cite{DVF,ST4} is defined by 
\ba
{\cal V}(x, \theta) \A = \A C(x) + \bar\theta^i \Lambda^i(x) 
+ {1 \over 2} \bar\theta^i \theta^j M^{ij}(x) 
- {1 \over 2} \bar\theta^i \theta^i M^{jj}(x) 
+ {1 \over 4} \epsilon^{ij} \bar\theta^i \gamma_5 \theta^j \phi(x) 
\nonu
\A \A 
- {i \over 4} \epsilon^{ij} \bar\theta^i \gamma_a \theta^j v^a(x) 
- {1 \over 2} \bar\theta^i \theta^i \bar\theta^j \lambda^j(x) 
- {1 \over 8} \bar\theta^i \theta^i \bar\theta^j \theta^j D(x), 
\label{VSF}
\ea
where among component fields $\varphi_{\cal V}^I(x) = \{ C(x), \Lambda^i(x), M^{ij}(x), \cdots \}$ 
in the superfield (\ref{VSF}), we denote 
$(C, D)$ for two scalar fields, $(\Lambda^i, \lambda^i)$ for two doublet (Majorana) spinor fields, 
$\phi$ for a pseudo scalar field, $v^a$ for a vector field, 
and $M^{ij} = M^{(ij)}$ $\left(= {1 \over 2}(M^{ij} + M^{ji}) \right)$ 
for three scalar fields ($M^{ii} = \delta^{ij} M^{ij}$). 
The $d = 2$, $N = 2$ scalar superfields are expressed as 
\ba
\Phi^i(x, \theta) \A = \A B^i(x) + \bar\theta^i \chi(x) - \epsilon^{ij} \bar\theta^j \nu(x) 
- {1 \over 2} \bar\theta^j \theta^j F^i(x) + \bar\theta^i \theta^j F^j(x) 
- i \bar\theta^i \!\!\not\!\partial B^j(x) \theta^j 
\nonu
\A \A 
+ {i \over 2} \bar\theta^j \theta^j (\bar\theta^i \!\!\not\!\partial \chi(x) 
- \epsilon^{ik} \bar\theta^k \!\!\not\!\partial \nu(x)) 
+ {1 \over 8} \bar\theta^j \theta^j \bar\theta^k \theta^k \Box B^i(x), 
\label{SSF}
\ea
where among component fields $\varphi_\Phi^I(x) = \{ B^i(x), \chi^i(x), \nu^i(x), F^i(x) \}$ 
in the superfield (\ref{SSF}), we denote 
$B^i$ for doublet scalar fields, $(\chi, \nu)$ for two (Majorana) spinor fields 
and $F^i$ for doublet auxiliary scalar fields. 
The supertransformations of the gauge and scalar superfields with constant (Majorana) spinor parameters $\zeta^i$ 
are given as 
\be
\delta_\zeta {\cal V}(x, \theta) = \bar\zeta^i Q^i {\cal V}(x, \theta), 
\ \ \delta_\zeta \Phi^i(x, \theta) = \bar\zeta^j Q^j \Phi^i(x, \theta), 
\label{SFtransfn}
\ee
where $Q^i = {\partial \over \partial\bar\theta^i} + i \!\!\not\!\!\!\partial \theta^i$ 
are the generators of LSUSY, which determine LSUSY transformations of the component fields 
in the power series expansion with respect to $\theta^i$. 

The general $N = 2$ SUSY QED gauge action with SSB is written in terms of the general gauge 
and the scalar matter superfields (\ref{VSF}) and (\ref{SSF}) (for the massless case) as 
\be
L^{\rm gen.}_{N = 2{\rm SUSYQED}} 
= L_{{\cal V}{\rm kin}} + L_{{\cal V}{\rm FI}} 
+ L_{\Phi{\rm kin}} + L_e 
\label{SQEDaction}
\ee
with 
\ba
L_{{\cal V}{\rm kin}} 
\A = \A {1 \over 32} \left\{ \int d^2 \theta^i 
\ (\overline{D^i {\cal W}^{jk}} D^i {\cal W}^{jk} 
+ \overline{D^i {\cal W}_5^{jk}} D^i {\cal W}_5^{jk}) \right\}_{\theta^i = 0}, 
%\nonu
%\A = \A 
%- {1 \over 4} (F_{0ab})^2 
%+ {i \over 2} \bar\lambda_0^i \!\!\not\!\partial \lambda_0^i 
%+ {1 \over 2} (\partial_a A_0)^2 + {1 \over 2} (\partial_a \phi_0)^2 + {1 \over 2} D_0^2, 
\label{Vkin}
\\
L_{{\cal V}{\rm FI}} 
\A = \A {1 \over 2} \int d^4 \theta^i \ {\xi \over \kappa} {\cal V}, %= - {\xi \over \kappa} (D_0 - \Box C), 
\label{VFI}
\\
L_{\Phi{\rm kin}} + L_e 
\A = \A - {1 \over 16} \int d^4 \theta^i \ e^{-4e{\cal V}} (\Phi^j)^2, 
%\nonu
%\A = \A 
%{i \over 2} \bar\chi \!\!\not\!\partial \chi 
%+ {1 \over 2} (\partial_a B^i)^2 
%+ {i \over 2} \bar\nu \!\!\not\!\partial \nu 
%+ {1 \over 2} (F^i)^2 
%- {1 \over 4} \partial_a (B^i \partial^a B^i) 
%\nonu
%\A \A 
%+ e \ \bigg\{ i v_{0a} \bar\chi \gamma^a \nu 
%- \epsilon^{ij} v_0^a B^i \partial_a B^j 
%+ \bar\lambda_0^i \chi B^i 
%+ \epsilon^{ij} \bar\lambda_0^i \nu B^j 
%- {1 \over 2} D_0 (B^i)^2 
%\nonu
%\A \A 
%+ {1 \over 2} A_0 (\bar\chi \chi + \bar\nu \nu) 
%- \phi_0 \bar\chi \gamma_5 \nu + \cdots \bigg\} 
%\nonu
%\A \A 
%+ {1 \over 2} e^2 \{ (v_{0a}{}^2 - A_0^2 - \phi_0^2) (B^i)^2 + \cdots \} 
%\nonu
%\A \A 
%+ \cdots. 
\label{gauge}
\ea
where $L_{{\cal V}{\rm kin}}$, $L_{{\cal V}{\rm FI}}$, $L_{\Phi{\rm kin}}$ and $L_e$ 
are the kinetic terms for the vector supermultiplet, the Fayet-Iliopoulos (FI) $D$ term, 
the kinetic terms for the scalar matter supermultiplet and the gauge interaction terms, respectively. 
In Eq.(\ref{Vkin}) ${\cal W}^{ij}$ and ${\cal W}_5^{ij}$ are 
scalar and pseudo scalar superfields defined by 
\be
{\cal W}^{ij} = \bar D^i D^j {\cal V}, \ \ \ {\cal W}_5^{ij} = \bar D^i \gamma_5 D^j {\cal V} 
\ee
with the differential operators $D^i = {\partial \over \partial\bar\theta^i} - i \!\!\not\!\!\partial \theta^i$. 
In Eq.(\ref{VFI}) $\xi$ is an arbitrary dimensionless parameter 
and $\kappa$ is a constant with the dimension (mass)$^{-1}$, 
while in Eq.(\ref{gauge}) $e$ is a gauge coupling constant whose dimension is (mass)$^1$ in $d = 2$. 

The explicit component form of the $N = 2$ SUSY QED action (\ref{SQEDaction}), 
i.e. the actions from (\ref{Vkin}) to (\ref{gauge}), is 
\ba
L_{{\cal V}{\rm kin}} \A = \A 
- {1 \over 4} (F_{0ab})^2 
+ {i \over 2} \bar\lambda_0^i \!\!\not\!\partial \lambda_0^i 
+ {1 \over 2} (\partial_a A_0)^2 + {1 \over 2} (\partial_a \phi_0)^2 + {1 \over 2} D_0^2 
\equiv L^0_{{\cal V}{\rm kin}}, 
\label{Vkin-comp}
\\
L_{{\cal V}{\rm FI}} \A = \A - {\xi \over \kappa} (D_0 - \Box C) 
\equiv L^0_{{\cal V}{\rm FI}} + {\xi \over \kappa} \Box C, 
\label{VFI-comp}
\\
L_{\Phi{\rm kin}} \A = \A 
{i \over 2} \bar\chi \!\!\not\!\partial \chi 
+ {1 \over 2} (\partial_a B^i)^2 
+ {i \over 2} \bar\nu \!\!\not\!\partial \nu 
+ {1 \over 2} (F^i)^2 
- {1 \over 4} \partial_a (B^i \partial^a B^i) 
\nonu
\A \equiv \A L^0_{\Phi{\rm kin}} - {1 \over 4} \partial_a (B^i \partial^a B^i), 
\label{Skin-comp}
\\
L_e \A = \A 
e \ \bigg\{ i v_{0a} \bar\chi \gamma^a \nu 
- \epsilon^{ij} v_0^a B^i \partial_a B^j 
+ \bar\lambda_0^i \chi B^i 
+ \epsilon^{ij} \bar\lambda_0^i \nu B^j 
- {1 \over 2} D_0 (B^i)^2 
\nonu
\A \A 
+ {1 \over 2} A_0 (\bar\chi \chi + \bar\nu \nu) 
- \phi_0 \bar\chi \gamma_5 \nu + \cdots \bigg\} 
\nonu
\A \A 
+ {1 \over 2} e^2 \{ (v_{0a}{}^2 - A_0^2 - \phi_0^2) (B^i)^2 + \cdots \} 
+ \cdots, 
\nonu
\A \equiv \A L^0_e + \cdots, 
\label{gauge-comp}
\ea
where gauge invariant quantities \cite{lin-ST4a,WB} are denoted by 
\be
\{ A_0, \phi_0, F_{0ab}, \lambda_0^i, D_0 \} 
\equiv \{ M^{ii}, \phi, F_{ab}, \lambda^i + i \!\!\not\!\partial \Lambda^i, D + \Box C \} 
\label{gauge-inv}
\ee
with $F_{0ab} = \partial_a v_{0b} - \partial_b v_{0a}$ and $F_{ab} = \partial_a v_b - \partial_b v_a$, 
which are invariant ($v_{0a} = v_a$ transforms as an Abelian gauge field) 
under a SUSY generalized gauge transformation, $\delta_g {\cal V} = \Lambda^1 + \alpha \Lambda^2$ \cite{WB,ST4}, 
with an arbitrary real parameter $\alpha$ and generalized gauge parameters $\Lambda^i$ 
in the form of the $N = 2$ scalar superfields. 
The component fields $\varphi_{{\cal V}0}^I = \{ A_0, \phi_0, v_{0a}, \lambda_0^i, D_0 \}$ in Eq.(\ref{gauge-inv}) 
correspond to the degrees of freedom (d.o.f.) for a minimal off-shell vector supermultiplet. 

In Eqs. from (\ref{Vkin-comp}) to (\ref{gauge-comp}) $L^0_{{\cal V}{\rm kin}}$, $L^0_{{\cal V}{\rm FI}}$, 
$L^0_{\Phi{\rm kin}}$ and $L^0_e$ are defined as the actions which are expressed 
in terms of only the component fields $\varphi_{{\cal V}0}^I$ and $\varphi_\Phi^I$, 
while the ellipses in Eq.(\ref{gauge-comp}) mean the terms depending explicitly 
on the redundant auxiliary fields $\{ C, \Lambda, M^{ij} (i \not= j) \}$ in the general gauge superfield 
and higher order terms of $e^n \ (n \ge 3)$. 
The $N = 2$ SUSY QED action (\ref{SQEDaction}) in the Wess-Zumino (WZ) gauge \cite{WB,ST4} gives 
the minimal action for the minimal off-shell vector supermultiplet with the arbitrary $e$. 
%Note that the field equation for $D$ (or $D_0$) through the FI $D$ term 
%induces the vev $<D> (= <D_0>) = {\xi \over \kappa}$ indicating the SSB. 

\section{Linearization of $N = 2$ NLSUSY in $d = 2$}

In order to discuss the general NL/L SUSY relation for the $N = 2$ SUSY QED theory in $d = 2$, 
%without imposing a priori any special gauge conditions, 
let us show in the linearization of $N = 2$ NLSUSY in $d = 2$ 
the SUSY invariant constraints and the subsequent SUSY invariant relations in the most general form. 
They are obtained by introducing NG fermions $\psi^i$ 
%with their supertransformations 
%(on a hypersurface defined by $\theta^i = \kappa \psi^i$ in the superspace), i.e. 
under NLSUSY transformations \cite{VA}, 
\be
\delta_\zeta \psi^i = {1 \over \kappa} \zeta^i 
- i \kappa \bar\zeta^j \gamma^a \psi^j \partial_a \psi^i. 
\label{NLSUSY}
\ee
The $N = 2$ NLSUSY transformations (\ref{NLSUSY}) make the following action invariant; 
namely, the $N = 2$ NLSUSY action is 
\be
L_{N = 2{\rm NLSUSY}} = - {1 \over {2 \kappa^2}} \ \vert w \vert, 
\label{NLSUSYaction}
\ee
where $\vert w \vert$ is the determinant \cite{VA} describing the dynamics of (massless) $\psi^i$, 
i.e. in $d = 2$, 
\be
\vert w \vert = \det(w^a{}_b) = \det(\delta^a_b + t^a{}_b) 
= 1 + t^a{}_a + {1 \over 2!}(t^a{}_a t^b{}_b - t^a{}_b t^b{}_a) 
\ee
with $t^a{}_b = - i \kappa^2 \bar\psi^i \gamma^a \partial_b \psi^i$ 
($\kappa^{-2} \sim {\Lambda \over G}$ in NLSUSY GR \cite{KS3}). 

The SUSY invariant relations which describe all the component fields 
in the $N = 2$ SUSY QED theory in terms of the NG fermions $\psi^i$ 
are systematically obtained by considering the superfields on specific superspace coordinates \cite{IK,UZ} 
shifted with a parameter $\zeta^i = - \kappa \psi^i$, 
which are denoted by $(x'^a, \theta_\alpha'^i)$, 
\ba
\A \A 
x'^a = x^a + i \kappa \bar\theta^i \gamma^a \psi^i, 
\nonu
\A \A 
\theta'^i = \theta^i - \kappa \psi^i. 
\ea
%
%where the origin of $(x'^a, \theta_\alpha'^i)$ 
%corresponds to the hypersurface $\theta^i = \kappa \psi^i$ for NLSUSY. 
Indeed, we define the $N = 2$ general gauge and the $N = 2$ scalar matter superfields 
on $(x'^a, \theta'^i)$ as 
\be
{\cal V}(x', \theta') \equiv \tilde{\cal V}(x, \theta, \psi(x)), 
\ \ \Phi^i(x', \theta') \equiv \tilde \Phi^i(x, \theta, \psi(x)), 
\label{SFpsi}
\ee
and their expansions around $(x^a, \theta^i)$ which terminate at ${\cal O}(\theta^4)$ are 
\ba
\tilde{\cal V}(x, \theta, \psi(x)) 
\A = \A \tilde C(x) + \bar\theta^i \tilde\Lambda^i(x) 
+ {1 \over 2} \bar\theta^i \theta^j \tilde M^{ij}(x) 
- {1 \over 2} \bar\theta^i \theta^i \tilde M^{jj}(x) 
+ {1 \over 4} \epsilon^{ij} \bar\theta^i \gamma_5 \theta^j \tilde\phi(x) 
\nonu
\A \A 
- {i \over 4} \epsilon^{ij} \bar\theta^i \gamma_a \theta^j \tilde v^a(x) 
- {1 \over 2} \bar\theta^i \theta^i \bar\theta^j \tilde\lambda^j(x) 
- {1 \over 8} \bar\theta^i \theta^i \bar\theta^j \theta^j \tilde D(x), 
\label{tVSF} \\
\tilde \Phi^i(x, \theta, \psi(x)) 
\A = \A \tilde B^i(x) + \bar\theta^i \tilde \chi(x) - \epsilon^{ij} \bar\theta^j \tilde \nu(x) 
- {1 \over 2} \bar\theta^j \theta^j \tilde F^i(x) + \bar\theta^i \theta^j \tilde F^j(x) + \cdots. 
\label{tSSF}
\ea
In Eqs.(\ref{tVSF}) and (\ref{tSSF}) the component fields 
$\tilde\varphi_{\cal V}^I(x) = \{ \tilde C(x), \tilde\Lambda^i(x), \cdots \}$ 
and $\tilde\varphi_\Phi^I(x) = \{ \tilde B^i(x), \tilde\chi(x), \cdots \}$ 
are expressed in terms of the component fields $\varphi_{\cal V}^I(x)$ and $\varphi_\Phi^I(x)$ 
in Eqs.(\ref{VSF}) and (\ref{SSF}) and the NG fermions $\psi^i$ \cite{lin-ST2,lin-ST4b}. 

According to the supertransformations (\ref{SFtransfn}) and (\ref{NLSUSY}), 
the superfields (\ref{SFpsi}) transform homogeneously \cite{IK,UZ} as 
\be
\delta_\zeta \tilde{\cal V}(x, \theta, \psi(x)) = \xi^a \partial_a \tilde{\cal V}(x, \theta, \psi(x)), 
\ \ \ \delta_\zeta \tilde \Phi^i(x, \theta, \psi(x)) = \xi^a \partial_a \tilde \Phi^i(x, \theta, \psi(x)) 
\ee
with $\xi^a = i \kappa \bar\psi^i \gamma^a \zeta^i$, 
which mean the components $\tilde\varphi_{\cal V}^I(x)$ 
and $\tilde\varphi_\Phi^I(x)$ do not transform each other, respectively. 
Therefore, the following conditions, i.e. the SUSY invariant constraints eliminating the other d.o.f. 
than $\varphi_{\cal V}^I(x)$, $\varphi_\Phi^I(x)$ and $\psi^i$, can be imposed, 
\ba
\A \A 
\tilde\varphi_{\cal V}^I(x) = {\rm constant}, 
\label{SUSYconst-VSF}
\\
\A \A 
\tilde\varphi_\Phi^I(x) = {\rm constant}, 
\label{SUSYconst-SSF}
\ea
which are invariant (conserved quantities) under the supertransformations (\ref{SFtransfn}) and (\ref{NLSUSY}). 

The constraints (\ref{SUSYconst-VSF}) and (\ref{SUSYconst-SSF}) 
are written in the most general form as follows; 
\ba
\A \A 
\tilde C = \xi_c, \ \ \tilde\Lambda^i = \xi_\Lambda^i, 
\ \ \tilde M^{ij} = \xi_M^{ij}, \ \ \tilde\phi = \xi_\phi, 
\ \ \tilde v^a = \xi_v^a, \ \ \tilde\lambda^i = \xi_\lambda^i, 
\ \ \tilde D = {\xi \over \kappa}, 
\label{SUSYconst-VSF1}
\\
\A \A 
\tilde B^i = \xi_B^i, \ \ \tilde\chi = \xi_\chi, \ \ \tilde\nu = \xi_\nu, 
\ \ \tilde F^i = {\xi^i \over \kappa}, 
\label{SUSYconst-SSF1}
\ea
where the mass dimensions of constants (or constant spinors) in $d = 2$ 
are defined by ($-1$, ${1 \over 2}$, $0$, $0$, $0$, $-{1 \over 2}$) 
for ($\xi_c$, $\xi_\Lambda^i$, $\xi_M^{ij}$, $\xi_\phi$, $\xi_v^a$, $\xi_\lambda^i$), 
($0$, $-{1 \over 2}$, $-{1 \over 2}$) for ($\xi_B^i$, $\xi_\chi$, $\xi_\nu$) and $0$ for $\xi^i$ for convenience. 
The general SUSY invariant constraints (\ref{SUSYconst-VSF1}) and (\ref{SUSYconst-SSF1}) 
can be solved with respect to the component fields $\varphi_{\cal V}^I$ and $\varphi_\Phi^I$ 
in terms of $\psi^i$; namely, the SUSY invariant relations $\varphi_{\cal V}^I = \varphi_{\cal V}^I(\psi)$ 
are calculated systematically and straightforwardly as 
\ba
C \A = \A \xi_c + \kappa \bar\psi^i \xi_\Lambda^i 
+ {1 \over 2} \kappa^2 (\xi_M^{ij} \bar\psi^i \psi^j - \xi_M^{ii} \bar\psi^j \psi^j) 
+ {1 \over 4} \xi_\phi \kappa^2 \epsilon^{ij} \bar\psi^i \gamma_5 \psi^j 
- {i \over 4} \xi_v^a \kappa^2 \epsilon^{ij} \bar\psi^i \gamma_a \psi^j 
\nonu
\A \A 
- {1 \over 2} \kappa^3 \bar\psi^i \psi^i \bar\psi^j \xi_\lambda^j 
- {1 \over 8} \xi \kappa^3 \bar\psi^i \psi^i \bar\psi^j \psi^j, 
\nonu
\Lambda^i \A = \A \xi_\Lambda^i 
+ \kappa (\xi_M^{ij} \psi^j - \xi_M^{jj} \psi^i) 
+ {1 \over 2} \xi_\phi \kappa \epsilon^{ij} \gamma_5 \psi^j 
- {i \over 2} \xi_v^a \kappa \epsilon^{ij} \gamma_a \psi^j 
\nonu
\A \A 
- {1 \over 2} \xi_\lambda^i \kappa^2 \bar\psi^j \psi^j 
+ {1 \over 2} \kappa^2 
(\psi^j \bar\psi^i \xi_\lambda^j 
- \gamma_5 \psi^j \bar\psi^i \gamma_5 \xi_\lambda^j 
- \gamma_a \psi^j \bar\psi^i \gamma^a \xi_\lambda^j) 
\nonu
\A \A 
- {1 \over 2} \xi \kappa^2 \psi^i \bar\psi^j \psi^j 
- i \kappa \!\!\not\!\partial C(\psi) \psi^i, 
\nonu
M^{ij} \A = \A \xi_M^{ij} 
+ \kappa \bar\psi^{(i} \xi_\lambda^{j)} 
+{1 \over 2} \xi \kappa \bar\psi^i \psi^j 
+ i \kappa \epsilon^{(i \vert k \vert} \epsilon^{j)l} \bar\psi^k \!\!\not\!\partial \Lambda^l(\psi) 
- {1 \over 2} \kappa^2 \epsilon^{ik} \epsilon^{jl} \bar\psi^k \psi^l \Box C(\psi), 
\nonu
\phi \A = \A \xi_\phi 
- \kappa \epsilon^{ij} \bar\psi^i \gamma_5 \xi_\lambda^j 
- {1 \over 2} \xi \kappa \epsilon^{ij} \bar\psi^i \gamma_5 \psi^j 
- i \kappa \epsilon^{ij} \bar\psi^i \gamma_5 \!\!\not\!\partial \Lambda^j(\psi) 
+ {1 \over 2} \kappa^2 \epsilon^{ij} \bar\psi^i \gamma_5 \psi^j \Box C(\psi), 
\nonu
v^a \A = \A \xi_v^a 
- i \kappa \epsilon^{ij} \bar\psi^i \gamma^a \xi_\lambda^j 
- {i \over 2} \xi \kappa \epsilon^{ij} \bar\psi^i \gamma^a \psi^j 
- \kappa \epsilon^{ij} \bar\psi^i \!\!\not\!\partial \gamma^a \Lambda^j(\psi) 
+ {i \over 2} \kappa^2 \epsilon^{ij} \bar\psi^i \gamma^a \psi^j \Box C(\psi) 
\nonu
\A \A 
- i \kappa^2 \epsilon^{ij} \bar\psi^i \gamma^b \psi^j \partial^a \partial_b C(\psi), 
\nonu
\lambda^i \A = \A \xi_\Lambda^i 
+ \xi \psi^i - i \kappa \!\!\not\!\partial M^{ij}(\psi) \psi^j 
+ {i \over 2} \kappa \epsilon^{ab} \epsilon^{ij} \gamma_a \psi^j \partial_b \phi(\psi) 
\nonu
\A \A 
- {1 \over 2} \kappa \epsilon^{ij} \left\{ \psi^j \partial_a v^a(\psi) 
- {1 \over 2} \epsilon^{ab} \gamma_5 \psi^j F_{ab}(\psi) \right\} 
\nonu
\A \A
- {1 \over 2} \kappa^2 \{ \Box \Lambda^i(\psi) \bar\psi^j \psi^j - \Box \Lambda^j(\psi) \bar\psi^i \psi^j 
- \gamma_5 \Box \Lambda^j(\psi) \bar\psi^i \gamma_5 \psi^j 
\nonu
\A \A 
- \gamma_a \Box \Lambda^j(\psi) \bar\psi^i \gamma^a \psi^j 
+ 2 \!\!\not\!\partial \partial_a \Lambda^j(\psi) \bar\psi^i \gamma^a \psi^j \} 
- {i \over 2} \kappa^3 \!\!\not\!\partial \Box C(\psi) \psi^i \bar\psi^j \psi^j, 
\nonu
D \A = \A {\xi \over \kappa} - i \kappa \bar\psi^i \!\!\not\!\partial \lambda^i(\psi) 
\nonu
\A \A 
+ {1 \over 2} \kappa^2 \left\{ \bar\psi^i \psi^j \Box M^{ij}(\psi) 
- {1 \over 2} \epsilon^{ij} \bar\psi^i \gamma_5 \psi^j \Box \phi(\psi) \right. 
\nonu
\A \A 
\left. 
+ {i \over 2} \epsilon^{ij} \bar\psi^i \gamma_a \psi^j \Box v^a(\psi) 
- i \epsilon^{ij} \bar\psi^i \gamma_a \psi^j \partial_a \partial_b v^b(\psi) \right\} 
\nonu
\A \A
- {i \over 2} \kappa^3 \bar\psi^i \psi^i \bar\psi^j \!\!\not\!\partial \Box \Lambda^j(\psi) 
+ {1 \over 8} \kappa^4 \bar\psi^i \psi^i \bar\psi^j \psi^j \Box^2 C(\psi), 
\label{SUSYrelation-VSF}
\ea
while the SUSY invariant relations $\varphi_\Phi^I = \varphi_\Phi^I(\psi)$ are 
\ba
B^i \A = \A \xi_B^i + \kappa (\bar\psi^i \xi_\chi - \epsilon^{ij} \bar\psi^j \xi_\nu) 
- {1 \over 2} \kappa^2 \{ \bar\psi^j \psi^j F^i(\psi) - 2 \bar\psi^i \psi^j F^j(\psi) 
+ 2 i \bar\psi^i \!\!\not\!\partial B^j(\psi) \psi^j \} 
\nonu
\A \A 
- i \kappa^3 \bar\psi^j \psi^j \{ \bar\psi^i \!\!\not\!\partial \chi(\psi) 
- \epsilon^{ik} \bar\psi^k \!\!\not\!\partial \nu(\psi) \} 
+ {3 \over 8} \kappa^4 \bar\psi^j \psi^j \bar\psi^k \psi^k \Box B^i(\psi), 
\nonu
\chi \A = \A \xi_\chi + \kappa \{ \psi^i F^i(\psi) - i \!\!\not\!\partial B^i(\psi) \psi^i \} 
\nonu
\A \A 
- {i \over 2} \kappa^2 [ \not\!\partial \chi(\psi) \bar\psi^i \psi^i 
- \epsilon^{ij} \{ \psi^i \bar\psi^j \!\!\not\!\partial \nu(\psi) 
- \gamma^a \psi^i \bar\psi^j \partial_a \nu(\psi) \} ] 
\nonu
\A \A 
+ {1 \over 2} \kappa^3 \psi^i \bar\psi^j \psi^j \Box B^i(\psi) 
+ {i \over 2} \kappa^3 \!\!\not\!\partial F^i(\psi) \psi^i \bar\psi^j \psi^j 
+ {1 \over 8} \kappa^4 \Box \chi(\psi) \bar\psi^i \psi^i \bar\psi^j \psi^j, 
\nonu
\nu \A = \A \xi_\nu - \kappa \epsilon^{ij} \{ \psi^i F^j(\psi) - i \!\!\not\!\partial B^i(\psi) \psi^j \} 
\nonu
\A \A 
- {i \over 2} \kappa^2 [ \not\!\partial \nu(\psi) \bar\psi^i \psi^i 
+ \epsilon^{ij} \{ \psi^i \bar\psi^j \!\!\not\!\partial \chi(\psi) 
- \gamma^a \psi^i \bar\psi^j \partial_a \chi(\psi) \} ] 
\nonu
\A \A 
+ {1 \over 2} \kappa^3 \epsilon^{ij} \psi^i \bar\psi^k \psi^k \Box B^j(\psi) 
+ {i \over 2} \kappa^3 \epsilon^{ij} \!\!\not\!\partial F^i(\psi) \psi^j \bar\psi^k \psi^k 
+ {1 \over 8} \kappa^4 \Box \nu(\psi) \bar\psi^i \psi^i \bar\psi^j \psi^j, 
\nonu
F^i \A = \A {\xi^i \over \kappa} - i \kappa \{ \bar\psi^i \!\!\not\!\partial \chi(\psi) 
+ \epsilon^{ij} \bar\psi^j \!\!\not\!\partial \nu(\psi) \} 
\nonu
\A \A 
- {1 \over 2} \kappa^2 \bar\psi^j \psi^j \Box B^i(\psi) + \kappa^2 \bar\psi^i \psi^j \Box B^j(\psi) 
+ i \kappa^2 \bar\psi^i \!\!\not\!\partial F^j(\psi) \psi^j 
\nonu
\A \A 
+ {1 \over 2} \kappa^3 \bar\psi^j \psi^j \{ \bar\psi^i \Box \chi(\psi) + \epsilon^{ik} \bar\psi^k \Box \nu(\psi) \} 
- {1 \over 8} \kappa^4 \bar\psi^j \psi^j \bar\psi^k \psi^k \Box F^i(\psi). 
\label{SUSYrelation-SSF}
\ea

For simplicity of arguments in NL/L SUSY relation, 
we reduce the above SUSY invariant constraints and the SUSY invariant relations, 
i.e. the massless eigenstates in terms of $\psi^i$, to more simple (but nontrivial and general) expressions. 
Since in Eqs.(\ref{SUSYrelation-VSF}) and (\ref{SUSYrelation-SSF}) 
the constants (the vevs) which do not couple to $\psi^i$ are only $\xi_c$ and $\xi_B^i$, we put 
\be
\xi_\Lambda^i = \xi_M^{ij} = \xi_\phi = \xi_v^a = \xi_\lambda^i = 0, 
\ \ \xi_\chi = \xi_\nu = 0. 
\ee
except for $\xi$ and $\xi^i$ which are the fundamental constants in the simplest 
and nontrivial NL/L SUSY relation for the $N = 2$ SUSY QED theory with the SSB \cite{lin-ST3,lin-ST4b}. 
Further we put 
\be
\xi_B^i = 0, 
\ee
because we would like to attribute straightforwardly the $N = 2$ SUSY QED action (\ref{SQEDaction}) 
to the $N = 2$ NLSUSY action (\ref{NLSUSYaction}) up to a normalization factor 
when the SUSY invariant relations are substituted into Eq.(\ref{SQEDaction}). 
Then, the SUSY invariant constraints (\ref{SUSYconst-VSF1}) and (\ref{SUSYconst-SSF1}) become 
\ba
\A \A 
\tilde C = \xi_c, \ \ \tilde\Lambda^i = \tilde M^{ij} = \tilde\phi = \tilde v^a = \tilde\lambda^i = 0, 
\ \ \tilde D = {\xi \over \kappa}, 
\label{SUSYconst-VSF2}
\\
\A \A 
\tilde B^i = \tilde\chi = \tilde\nu = 0, \ \ \ \tilde F^i = {\xi^i \over \kappa}, 
\label{SUSYconst-SSF2}
\ea
and the SUSY invariant relations (\ref{SUSYrelation-VSF}) and (\ref{SUSYrelation-SSF}) reduce to 
\ba
C \A = \A \xi_c - {1 \over 8} \xi \kappa^3 \bar\psi^i \psi^i \bar\psi^j \psi^j \vert w \vert, 
\nonu
\Lambda^i \A = \A - {1 \over 2} \xi \kappa^2 
\psi^i \bar\psi^j \psi^j \vert w \vert, %(1 - i \kappa^2 \bar\psi^k \!\!\not\!\partial \psi^k), 
\nonu
M^{ij} \A = \A {1 \over 2} \xi \kappa \bar\psi^i \psi^j \vert w \vert, 
%\left( 1 - i \kappa^2 \bar\psi^k \!\!\not\!\partial \psi^k 
%- {1 \over 2} \kappa^4 \epsilon^{ab} \bar\psi^k \psi^l 
%\partial_a \bar\psi^k \gamma_5 \partial_b \psi^l \right), 
\nonu
\phi \A = \A - {1 \over 2} \xi \kappa \epsilon^{ij} \bar\psi^i \gamma_5 \psi^j \vert w \vert, 
%\left( 1 - i \kappa^2 \bar\psi^k \!\!\not\!\partial \psi^k 
%- {1 \over 2} \kappa^4 \epsilon^{ab} \bar\psi^k \gamma_5 \psi^l 
%\partial_a \bar\psi^k \partial_b \psi^l \right), 
%
\nonu
v^a \A = \A - {i \over 2} \xi \kappa \epsilon^{ij} \bar\psi^i \gamma^a \psi^j \vert w \vert, 
%(1 - i \kappa^2 \bar\psi^k \!\!\not\!\partial \psi^k), 
\nonu
\lambda^i \A = \A \xi \psi^i \vert w \vert, 
\nonu
D \A = \A {\xi \over \kappa} \vert w \vert, 
\label{SUSYrelation-VSF1}
\end{eqnarray}
and 
\ba
\chi \A = \A \xi^i \left[ \psi^i \vert w \vert
+ {i \over 2} \kappa^2 \partial_a 
( \gamma^a \psi^i \bar\psi^j \psi^j \vert w \vert 
%(1 - i \kappa^2 \bar\psi^k \!\!\not\!\partial \psi^k) 
) \right], 
\nonu
B^i \A = \A - \kappa \left( {1 \over 2} \xi^i \bar\psi^j \psi^j 
- \xi^j \bar\psi^i \psi^j \right) \vert w \vert, 
\nonu
\nu \A = \A \xi^i \epsilon^{ij} \left[ \psi^j \vert w \vert 
+ {i \over 2} \kappa^2 \partial_a 
( \gamma^a \psi^j \bar\psi^k \psi^k \vert w \vert 
%(1 - i \kappa^2 \bar\psi^l \!\!\not\!\partial \psi^l) 
) \right], 
\nonu
F^i \A = \A {1 \over \kappa} \xi^i \left\{ \vert w \vert 
+ {1 \over 8} \kappa^3 
\Box ( \bar\psi^j \psi^j \bar\psi^k \psi^k \vert w \vert ) 
\right\} 
\nonu
\A \A 
- i \kappa \xi^j \partial_a ( \bar\psi^i \gamma^a \psi^j \vert w \vert ), 
%- {1 \over 4} e \kappa^2 \xi \xi^i \bar\psi^j \psi^j \bar\psi^k \psi^k. 
\label{SUSYrelation-SSF1}
\end{eqnarray}
which are written in the form containing some vanishing terms due to $(\psi^i)^5 \equiv 0$.

\section{NL/L SUSY relation for $N = 2$ SUSY QED in $d = 2$}

In this section we discuss the relation between the $N = 2$ SUSY QED action (\ref{SQEDaction}) 
and the $N = 2$ NLSUSY action (\ref{NLSUSYaction}). 
Substituting the reduced (but general) SUSY invariant relations (\ref{SUSYrelation-VSF1}) 
and (\ref{SUSYrelation-SSF1}) into Eqs.(\ref{Vkin}), (\ref{VFI}) and (\ref{gauge}) 
gives the relations among the actions as follows; 
\ba
L_{{\cal V}{\rm kin}}(\psi) \A = \A - \xi^2 L_{N = 2{\rm NLSUSY}}, 
\nonu
L_{{\cal V}{\rm FI}}(\psi) \A = \A 2 \xi^2 L_{N = 2{\rm NLSUSY}}, 
\nonu
(L_{\Phi{\rm kin}} + L_e)(\psi) \A = \A - (\xi^i)^2 e^{-4 e \xi_c} L_{N = 2{\rm NLSUSY}}. 
\label{NL-LSUSY}
\ea
These results can be obtained systematically by changing the integration variables 
in the actions (\ref{Vkin}), (\ref{VFI}) and (\ref{gauge}) 
from $(x, \theta^i)$ to $(x', \theta'^i)$ under the SUSY invariant constraints 
(\ref{SUSYconst-VSF2}) and (\ref{SUSYconst-SSF2}) (see, for example, \cite{lin-ST2}). 
Therefore, from Eq.(\ref{NL-LSUSY}) we obtain a general NL/L SUSY relation 
for the $N = 2$ SUSY QED theory in $d = 2$ as 
\be
f(\xi, \xi^i, \xi_c, e) \ L_{N = 2{\rm NLSUSY}} = L^{\rm gen.}_{N = 2{\rm SUSYQED}} 
\label{NL-LSUSYgen}
\ee
with a normalization factor $f(\xi, \xi^i, \xi_c, e)$ defined by 
\be
f(\xi, \xi^i, \xi_c, e) = \xi^2 - (\xi^i)^2 e^{-4 e \xi_c}. 
\label{normalization}
\ee

By regarding the NLSUSY GR theory in the SGM scenario as the fundamental theory of space-time and matter, 
$L^{\rm gen.}_{N = 2{\rm SUSYQED}}$ is attributed to $L_{N = 2{\rm NLSUSY}}$ which is the cosmological term 
of SGM (NLSUSY GR) action for flat space-time, i.e. we put 
\be
f(\xi, \xi^i, \xi_c, e) = 1. 
\label{normalization1}
\ee
Remarkably, the condition (\ref{normalization1}) gives the gauge coupling constant $e$ 
in terms of $\xi$, $\xi^i$ and $\xi_c$ as 
\be
e = {1 \over 4\xi_c} \ln X, 
\ \ \ X = {(\xi^i)^2 \over {\xi^2 - 1}}. 
\label{gcoupling}
\ee

Finally we just mention the relation between the general $N = 2$ SUSY QED action (\ref{SQEDaction}) 
and the {\it minimal} one for the minimal off-shell vector supermultiplet 
in the NL/L SUSY relation (\ref{NL-LSUSYgen}) with the normalization condition (\ref{normalization1}). 
The minimal actions $L^0_{{\cal V}{\rm kin}}$, $L^0_{{\cal V}{\rm FI}}$, 
$L^0_{\Phi{\rm kin}}$ and $L^0_e$ defined in Eqs. from (\ref{Vkin-comp}) to (\ref{gauge-comp}) 
is related to the general actions of Eq.(\ref{NL-LSUSY}) 
for the $N = 2$ SUSY QED theory in NL/L SUSY relation as 
\ba
L_{{\cal V}{\rm kin}}(\psi) \A = \A L^0_{{\cal V}{\rm kin}}(\psi) = - \xi^2 L_{N = 2{\rm NLSUSY}}, 
\nonu
L_{{\cal V}{\rm FI}}(\psi) \A = \A L^0_{{\cal V}{\rm FI}}(\psi) + [{\rm tot.\ der.\ terms}] 
= 2 \xi^2 L_{N = 2{\rm NLSUSY}}, 
\nonu
(L_{\Phi{\rm kin}} + L_e)(\psi) 
\A = \A (e^{-4 e \xi_c} L^0_{\Phi{\rm kin}}\vert_{F \rightarrow F'} + L^0_e)(\psi) 
%- {1 \over 4} e \kappa \xi (\xi^i)^2 \bar\psi^j \psi^j\bar\psi^k \psi^k 
%\nonu
%\A \A 
+ [{\rm tot.\ der.\ terms}] 
\nonu
\A = \A - (\xi^i)^2 e^{-4 e \xi_c} L_{N = 2{\rm NLSUSY}}, 
\label{NL-LSUSY0}
\ea
where $L^0_e(\psi) = {1 \over 4} e \kappa \xi (\xi^i)^2 \bar\psi^j \psi^j\bar\psi^k \psi^k$ 
%
%\ba
%\A \A 
%L^0_{{\cal V}{\rm kin}}(\psi) = L_{{\cal V}{\rm kin}}(\psi) = - \xi^2 L_{N = 2{\rm NLSUSY}}, 
%\nonu
%\A \A 
%L^0_{{\cal V}{\rm FI}}(\psi) = L_{{\cal V}{\rm FI}}(\psi) + [{\rm tot.\ der.\ terms}] 
%= 2 \xi^2 L_{N = 2{\rm NLSUSY}}, 
%\nonu
%\A \A 
%L^0_{\Phi{\rm kin}}(\psi) = L_{\Phi{\rm kin}}(\psi) + [{\rm tot.\ der.\ terms}] 
%= - (\xi^i)^2 L_{N = 2{\rm NLSUSY}}, 
%\nonu
%\A \A 
%L^0_e(\psi) = L_e(\psi) + {1 \over 4} e \kappa \xi (\xi^i)^2 \bar\psi^j \psi^j\bar\psi^k \psi^k 
%= {1 \over 4} e \kappa \xi (\xi^i)^2 \bar\psi^j \psi^j\bar\psi^k \psi^k, 
%\ea
%
and the SUSY invariant relations of the auxiliary fields $F^i$ in Eq.(\ref{SUSYrelation-SSF1}) 
have been changed (relaxed) by four NG-fermion self-interaction terms as 
\be
F'^i(\psi) = F^i(\psi) - {1 \over 4} e^{4 e \xi_c} e \kappa^2 \xi \xi^i \bar\psi^j \psi^j \bar\psi^k \psi^k. 
%\label{gen-F}
\ee
Obviously, the minimal $N = 2$ SUSY QED action for the minimal off-shell vector supermultiplet 
is included in the relations (\ref{NL-LSUSY0}) at the leading order of the factor $e^{-4 e \xi_c}$. 

It can be seen easily that the numerical factor $e^{-4 e \xi_c}$ in the relation (\ref{NL-LSUSY0}) 
is absorbed into the action by rescaling the whole scalar supermultiplet 
$\varphi_\Phi^I = \{ B^i, \chi^i, \nu^i,$ $F^i \}$ in the scalar superfields $\Phi^i$ by $e^{-2 e \xi_c}$ 
and by translating the auxiliary field $C$ by $\xi_c$ in the gauge action (\ref{gauge}) as 
\ba
\A \A 
\varphi_\Phi^I(\psi) \rightarrow \hat \varphi_\Phi^I(\psi) = e^{-2 \xi_C e} \varphi_\Phi^I(\psi), 
\nonu
\A \A 
C(\psi) \rightarrow \hat C(\psi) = - {1 \over 8} \xi \kappa^3 \bar\psi^i \psi^i \bar\psi^j \psi^j \vert w \vert. 
\label{rescale}
\ea
Indeed, in NL/L SUSY relation the actions (\ref{Vkin}), (\ref{VFI}) and (\ref{gauge}) 
in terms of the component fields $\{ \hat \varphi_\Phi^I, \hat C \}$ 
in addition to $\varphi_{{\cal V}0}^I = \hat \varphi_{{\cal V}0}^I$ in Eq.(\ref{gauge-inv}) become 
\ba
L_{{\cal V}{\rm kin}}(\psi) \A = \A \hat L_{{\cal V}{\rm kin}}(\psi) 
= \hat L^0_{{\cal V}{\rm kin}}(\psi) = - \xi^2 L_{N = 2{\rm NLSUSY}}, 
%\label{Vkin-hat}
\nonu
L_{{\cal V}{\rm FI}}(\psi) \A = \A \hat L_{{\cal V}{\rm FI}}(\psi) 
= \hat L^0_{{\cal V}{\rm FI}}(\psi) + [{\rm tot.\ der.\ terms}] 
\nonu
\A = \A 2 \xi^2 L_{N = 2{\rm NLSUSY}}, 
%\label{VFI-hat}
\nonu
(L_{\Phi{\rm kin}} + L_e)(\psi) \A = \A (\hat L_{\Phi{\rm kin}} + \hat L_e)(\psi) 
\nonu
\A = \A (\hat L^0_{\Phi{\rm kin}}\vert_{\hat F \rightarrow \hat F'} + \hat L^0_e)(\psi) 
+ [{\rm tot.\ der.\ terms}] 
\nonu
\A = \A - (\xi^i)^2 e^{-4 e \xi_c} L_{N = 2{\rm NLSUSY}}, 
\label{gauge-hat}
\ea
where $\hat L^0_e(\psi) 
= {1 \over 4} e \kappa \xi (\xi^i)^2 e^{-4 e \xi_c} \bar\psi^j \psi^j\bar\psi^k \psi^k$ and 
\be
\hat F'^i(\psi) =  e^{-2 e \xi_c} \left\{ F^i(\psi) 
- {1 \over 4}e \kappa^2 \xi \xi^i \bar\psi^j \psi^j \bar\psi^k \psi^k \right\}. 
%\label{gen-F}
\ee

Therefore, we obtain the ordinary $N = 2$ SUSY QED $U(1)$ gauge action 
for the minimal off-shell vector supermultiplet 
with the $U(1)$ gauge coupling constant (\ref{gcoupling}), i.e. 
\be
L_{N = 2{\rm NLSUSY}} = L^{\rm gen.}_{N = 2{\rm SUSYQED}} 
= \hat L^0_{N = 2{\rm LSUSYQED}} + [{\rm tot.\ der.\ terms}], 
\label{NL-LSUSYgen0}
\ee
where the minimal $N = 2$ SUSY QED $U(1)$ gauge action $\hat L^0_{N = 2{\rm LSUSYQED}}$ is defined by 
\be
\hat L^0_{N = 2{\rm LSUSYQED}} 
= \hat L^0_{{\cal V}{\rm kin}} + \hat L^0_{{\cal V}{\rm FI}} 
+ \hat L^0_{\Phi{\rm kin}}\vert_{\hat F \rightarrow \hat F'} + \hat L^0_e. 
\label{minSQEDaction}
\ee
Interestingly $e$ defined by the action (\ref{gauge}) depends upon the vevs of the auxiliary fields, i.e. 
the vacuum structures. 
Note that the bare $e$ is a free independent parameter, provided  $\xi_c = 0$ as in the case of adopting 
the WZ gauge throughout the arguments.

\section{Summary and discussion}

In this paper we have studied the NL/L SUSY relation for the $N = 2$ SUSY QED theory in $d = 2$ 
starting from the most general SUSY invariant constraints (\ref{SUSYconst-VSF1}) and (\ref{SUSYconst-SSF1}) 
and the subsequent SUSY invariant relations (\ref{SUSYrelation-VSF}) and (\ref{SUSYrelation-SSF}). 
After reducing those constraints and relations to more simple (but nontrivial and general) expressions 
of Eqs. from (\ref{SUSYconst-VSF2}) to (\ref{SUSYrelation-SSF1}), 
we have obtained the general NL/L SUSY relation (\ref{NL-LSUSYgen}) 
for the general $N = 2$ SUSY QED gauge action (\ref{SQEDaction}), 
which produces the normalization factor (\ref{normalization}) depending on the gauge coupling constant $e$. 
The fundamental notions of the NLSUSY GR theory in SGM scenario gives 
the normalization condition (\ref{normalization1}) and the relation 
between $e$ and the vevs (the constant terms in SUSY invariant relations) of the auxiliary scalar fields, 
$\xi$, $\xi^i$ and $\xi_c$ as in Eq.(\ref{gcoupling}). 

The minimal $N = 2$ LSUSY QED $U(1)$ gauge action (\ref{minSQEDaction}) 
has been also obtained from the general $N = 2$ SUSY QED action (\ref{SQEDaction}) 
related to the $N = 2$ NLSUSY action (\ref{NLSUSYaction}) 
as shown in the NL/L SUSY relation (\ref{NL-LSUSYgen0}). 
This has been achieved substantially by adopting the WZ gauge for LSUSY which can gauge away all auxiliary fields except $D_0$ 
in the vector supermultiplet and by rescaling the whole scalar supermultiplet in the gauge action (\ref{gauge}) 
by the constant numerical factor $e^{-2e \xi_c}$. 
However, the bare $U(1)$ gauge coupling constant $e$ in the minimal action (\ref{minSQEDaction}) 
is still defined as Eq.(\ref{gcoupling}) in terms of $\xi$, $\xi^i$ and $\xi_c$. 

The NLSUSY GR action $L_{\rm NLSUSYGR}(w)$ (in terms of a unified vierbein $w^a{}_\mu$) \cite{KS3} 
desribes geometrically the basic principle, 
i.e. the ultimate shape of nature is empty unstable space-time with the constant energy density 
(cosmological term) $\Lambda > 0$ and decays to (creates) spontaneously (quantum mechanically) 
ordinary Riemann space-time and spin-${1 \over 2}$ massless NG fermion superons matter 
with the potential (cosmological term) $V = \Lambda > 0$ depicted by the SGM action 
$L_{\rm SGM}(e, \psi)$ (in terms of the ordinary vierbein $e^a{}_\mu$ and the NG fermions $\psi^i$). 
We showed explicitly in asymptotic Riemann-flat ($e^a{}_\mu \rightarrow \delta^a{}_\mu$) space-time 
that the vacuum (the true minimum) of $L_{\rm SGM}(e, \psi)$ is $V = 0$, which is achieved 
when all local fields of the LSUSY multiplet of the familiar LSUSY gauge theory 
$L_{\rm LSUSY}(A, \lambda, v_a, \dots)$ are composed of NG-fermion superons 
dictated by the space-time $N$-extended NLSUSY symmetry, 
i.e. LSUSY is realized on the true vacuum as the composite eigenstates of SP symmetry. 

We can interpret the SM ($SU(5)$ GUT) in terms of superon picture of all particles 
and we obtain new remarkable insights into the proton decay (stable), neutrino oscillations, 
the origins of various mixing, CP-violation phase, etc. \cite{STS} 
by replacing the single line of the propagator of a particle in the Feynman diagrams of SM ($SU(5)$ GUT) 
by the multiple lines of superons. 

Our study may indicate that the general structure of the auxiliary fields for the general (gauge) superfield 
plays a crucial role in SUSY theory by determining not only the true vacuum through the SSB due to $D$ term 
but also the magnitude of the (bare) gauge coupling constant through the NL/L SUSY relation 
(i.e. the SUSY compositeness condition for all particles and the auxiliary fields as well), 
which is favorable to the SGM scenario for unity of nature. 
The strength of the bare gauge coupling constant may ought to be predicted, 
provided the fundamental theory is that of everything. 
The similar arguments in $d = 4$ for more general SUSY invariant constraints 
and for the large $N$ SUSY, especially $N = 4, 5$ are interesting and crucial.

\newpage

%%%%%%%  References  %%%%%%%%%%%%%%%%%%%%%%%%%%%%%%%%%%%%%%%
%
\newcommand{\NP}[1]{{\it Nucl.\ Phys.\ }{\bf #1}}
\newcommand{\PL}[1]{{\it Phys.\ Lett.\ }{\bf #1}}
\newcommand{\CMP}[1]{{\it Commun.\ Math.\ Phys.\ }{\bf #1}}
\newcommand{\MPL}[1]{{\it Mod.\ Phys.\ Lett.\ }{\bf #1}}
\newcommand{\IJMP}[1]{{\it Int.\ J. Mod.\ Phys.\ }{\bf #1}}
\newcommand{\PR}[1]{{\it Phys.\ Rev.\ }{\bf #1}}
\newcommand{\PRL}[1]{{\it Phys.\ Rev.\ Lett.\ }{\bf #1}}
\newcommand{\PTP}[1]{{\it Prog.\ Theor.\ Phys.\ }{\bf #1}}
\newcommand{\PTPS}[1]{{\it Prog.\ Theor.\ Phys.\ Suppl.\ }{\bf #1}}
\newcommand{\AP}[1]{{\it Ann.\ Phys.\ }{\bf #1}}


\begin{thebibliography}{100}

\bibitem{KS1}
K. Shima, {\it Z. Phys. C} {\bf 18} (1983) 25.

\bibitem{KS2}
K. Shima, {\it European Phys. J. C} {\bf 7} (1999) 341. 

\bibitem{KS3}
K. Shima, {\it Phys. Lett. B} {\bf 501} (2001) 237. 

\bibitem{VA}
D.V. Volkov and V.P. Akulov,  
%{\it JETP Lett.} {\bf 16} (1972) 438; 
{\it Phys. Lett. B} {\bf 46} (1973) 109. 

\bibitem{WZ}
J. Wess and B. Zumino, {\it Phys. Lett. B} {\bf 49} (1974) 52. 

\bibitem{GL}
Yu. A. Golfand and E.S. Likhtman, {\it JETP Lett.} {\bf 13} (1971) 323. 

%\bibitem{KS4}
%K. Shima, {\it Phys. Rev. D} {\bf 20} (1979) 574; {\it ibid} {\bf 15} (1977) 2165. 

\bibitem{ST1}
K. Shima and M. Tsuda, {\it Phys. Lett. B} {\bf 507} (2001) 260; 
{\it Class. Quant. Grav.} {\bf 19} (2002) 5101. 

%%%%%%%%%%%%%%%%%%%%

\bibitem{IK}
%E.A. Ivanov and A.A. Kapustnikov, 
%Relation between linear and nonlinear realizations of 
%supersymmetry, JINR Dubna Report No. E2-10765, 1977 (unpublished); \\
E.A. Ivanov and A.A. Kapustnikov, {\it J. Phys. A} {\bf 11} (1978) 2375; {\it J. Phys. G} {\bf 8} (1982) 167. 

\bibitem{Ro}
M. Ro\v{c}ek, {\it Phys. Rev. Lett.} {\bf 41} (1978) 451. 

\bibitem{UZ}
T. Uematsu and C.K. Zachos, {\it Nucl. Phys. B} {\bf 201} (1982) 250. 

\bibitem{STT}
K. Shima, Y. Tanii and M. Tsuda, {\it Phys. Lett. B} {\bf 546} (2002) 162; {\it ibid} {\bf 525} (2002) 183. 

\bibitem{lin-ST1}
K. Shima and M. Tsuda, {\it Phys. Lett. B} {\bf 641} (2006) 101. 

\bibitem{lin-ST2}
K. Shima and M. Tsuda, {\it Mod. Phys. Lett. A} {\bf 23} (2008) 3149. 

%%%%%%%%%%%%%%%%%%%%

\bibitem{lin-ST3}
K. Shima and M. Tsuda, {\it Mod. Phys. Lett. A} {\bf 22} (2007) 3027; {\it ibid} {\bf 22} (2007) 1085. 

\bibitem{lin-ST4a}
K. Shima and M. Tsuda, {\it Phys. Lett. B} {\bf 666} (2008) 410. 

\bibitem{lin-ST4b}
K. Shima and M. Tsuda, {\it Mod. Phys. Lett. A} {\bf 24} (2009) 185. 

%%%%%%%%%%%%%%%%%%%%

\bibitem{ST2}
K. Shima and M. Tsuda, {\it PoS HEP2005} (2006) 011. 

\bibitem{ST3}
K. Shima and M. Tsuda, {\it Phys. Lett. B} {\bf 645} (2007) 455. 

\bibitem{STL}
K. Shima, M. Tsuda and W. Lang, {\it Phys. Lett. B} {\bf 659} (2008) 741, 
Erratum {\it ibid} {\bf 660} (2008) 612, {\it ibid} {\bf 672} (2009) 413, 
arXiv:0710.1680 [hep-th]. 

\bibitem{STgc1}
K. Shima and M. Tsuda, 
Gauge coupling constant, compositeness and supersymmetry, arXiv:0810.5006 [hep-th]. 

\bibitem{STgc2}
K. Shima and M. Tsuda, {\it Fortsch. Phys.} {\bf 57} (2009) 698. 

\bibitem{WB}
J. Wess and J. Bagger, {\it Supersymmetry and Supergravity 
(Second Edition, Revised and Expanded)} 
(Princeton University Press, Princeton, New Jersey, 1992). 

\bibitem{DVF}
P. Di Vecchia and S. Ferrara, {\it Nucl. Phys. B} {\bf 130} (1977) 93. 

\bibitem{ST4}
K. Shima and M. Tsuda, {\it Mod. Phys. Lett. A} {\bf 23} (2008) 1167. 

\bibitem{STS}
K. Shima, M. Tsuda and M. Sawaguchi, {\it Int. J. Mod. Phys. E} {\bf 13} (2004) 539. 















\end{thebibliography}
\end{document}